\documentclass[12pt]{iopart}
\usepackage{iopams}
\usepackage{graphicx}

\newcommand{\lw}{\mathrm{lw}}
\newcommand{\tb}{\mathrm{tb}}

\newcommand{\dd}{\mathrm{d}}

\newcommand{\eq}[1]{(\ref{#1})}
\newcommand{\bun}{\hat{\mathbf{b}}}
\newcommand{\eun}{\hat{\mathbf{e}}}

\newcommand{\boldr}{\mathbf{r}}

\newcommand{\bv}{\mathbf{v}}

\newcommand{\bk}{\mathbf{k}}
\newcommand{\bR}{\mathbf{R}}

\newcommand{\bB}{\mathbf{B}}
\newcommand{\bE}{\mathbf{E}}

\newcommand{\matrixtop}[1]{\buildrel\leftrightarrow\over{#1}}
\newcommand{\matI}{\matrixtop{\mathbf{I}}}

\newcommand{\dotcross}{ \raise 0.65ex\hbox{${\scriptstyle {{_{\displaystyle \cdot}}\atop\times}}$} }
\newcommand{\crossdot}{ \raise 0.5ex\hbox{${\scriptstyle {{_\times}\atop{\displaystyle \cdot}}}$} }
\newcommand{\rhobf}{\mbox{\boldmath$\rho$}}

\newcommand{\kappabf}{\mbox{\boldmath$\kappa$}}

\newcommand{\sumsig}{ \raise -1.3ex\hbox{${{\displaystyle \sum}\atop{\scriptstyle \sigma}}$} }
\newcounter{appnumb}

\begin{document}
\title{Equivalence of two different approaches to global $\delta f$ gyrokinetic simulations}
\author{Felix I Parra and Michael Barnes}
\address{Rudolf Peierls Centre for Theoretical Physics, University of Oxford, Oxford, UK}
\address{Culham Centre for Fusion Energy, Abingdon, UK}
\eads{\mailto{f.parradiaz1@physics.ox.ac.uk}}

\begin{abstract}
A set of flux tube gyrokinetic equations that includes the effect of the spatial variation of the density, temperature and rotation gradients on the turbulence is derived. This new set of equations uses periodic boundary conditions. In the limit $l_\bot/L \ll 1$, where $l_\bot$ is the characteristic perpendicular length of turbulent structures and $L$ is the characteristic size of the device, this new set of flux tube gyrokinetic equations is shown to be equivalent to the traditional global $\delta f$ gyrokinetic equations to an order higher in $l_\bot/L$ than the usual flux tube formulations.
\end{abstract}

\submitto{\PPCF}
\maketitle

\section{Introduction} \label{sec:intro}
Tokamak and stellarator turbulence has characteristic lengths of the order of the ion gyroradius $\rho_i$ in the direction perpendicular to the magnetic field, and characteristic time scales of order $L/v_{ti}$, where $L$ is the characteristic length of the device \footnote{In this article, we do not assume that the inverse aspect ratio or the ratio $B_p/B$, where $B_p$ is the poloidal magnetic field and $B$ is the total magnetic field, are small. As a result, the characteristic length of the device along and across the magnetic field are considered of the same order.}, and $v_{ti}$ is the ion thermal speed \cite{mckee01}. Gyrokinetics \cite{catto78, frieman82} is the appropriate model for this type of turbulence because it permits averaging out the gyromotion, fast compared to the turbulence characteristic time, while keeping finite gyroradius effects. 

Because the perpendicular characteristic length of the turbulent eddies $l_\bot$ is of the order of the ion gyroradius, the typical perpendicular correlation length of the turbulence is smaller than the size of the device by a factor $\rho_\ast = \rho_i/L \ll 1$. It is then natural to assume that the turbulence characteristics depend only on the local values of density, temperature, electric field and magnetic field, and on their gradients. The flux tube formulation \cite{beer95} uses magnetic field aligned coordinates and periodic boundary conditions in the directions perpendicular to the magnetic field to give equations for the turbulence that only depend on the local background quantities. The justification for the periodic boundary conditions is statistical periodicity, i.e., beyond a perpendicular correlation length, the turbulence is statistically the same and uncorrelated, and as a result, we can simulate the turbulence in a box with periodic boundary conditions as long as the box is several correlation lengths across. This formulation has been implemented successfully in a number of codes \cite{dorland00, dannert05, watanabe06, peeters09c, numata10}. The fact that the results of these codes converge when resolution scans are performed suggests that the flux tube formulation is self-consistent, that is, it is the rigorous limit of the gyrokinetic equations for $l_\bot/L \sim \rho_\ast \rightarrow 0$.

The flux tube gyrokinetic formulation fails when the perpendicular size of the turbulent eddies $l_\bot$ is comparable to the size of the device, $l_\bot \sim L$. The surmise that flux tube codes may not be applicable to small machines in which $L$ is small led to the development of global $\delta f$ codes \cite{candy03, chen03, mcmillan08, goerler11} and full $f$ codes \cite{grandgirard06, xu07, chang08, heikkinen08} that do not assume statistical periodicity in the radial direction. The results obtained with these global codes tend to the solutions obtained with flux tube simulations when the size of the turbulent eddies is much smaller than the simulation domain \cite{goerler11, candy04}, confirming that the flux tube formulation is the correct limit for $l_\bot/L \sim \rho_\ast \rightarrow 0$.

In the derivation of the gyrokinetic equations, the eddy characteristic length is ordered as $l_\bot \sim \rho_i$ to include finite gyroradius effects in the equations. For this reason, it has been argued that devices with relatively large $\rho_\ast$ can only be modeled with global gyrokinetic codes, but this statement is not completely correct. Global codes can treat turbulence with $l_\bot \sim L$, but they are not particularly well suited for treating most of the finite $\rho_\ast$ effects. Flux tube gyrokinetic formulations are based on two independendent expansions: one in the ratio $\rho_\ast = v_{ti}/(L\Omega_i) = \rho_i/L \ll 1$ between the frequency of the turbulence $v_{ti}/L$ and the ion gyrofrequency $\Omega_i$, and one in the ratio $l_\bot/L \ll 1$ between the eddy perpendicular length and the size of the machine. Due to the expansion in $\rho_\ast$, it is possible to average over the fast gyromotion time scale. The expansion in $\rho_\ast$ is common to both global and flux tube codes, and finite $\rho_\ast$ effects can only be treated properly by keeping higher order terms in the gyrokinetic equation \cite{parra11a}, or by resorting to a full Vlasov formulation. The expansion in $l_\bot/L \ll 1$ is relaxed in global simulations. Thus, global codes are appropriate for $l_\bot \sim L \gg \rho_i$. The size of $l_\bot$ is connected to the size of the turbulent energy flux $Q^\tb$ which is in turn controlled by the energy injected into the machine. To illustrate the relation between $l_\bot$ and $Q^\tb$, we use the regime of Ion Temperature Gradient (ITG) turbulence far from marginal stability studied in \cite{barnes11b}. In this regime, the turbulent energy flux is
\begin{equation} \label{eq:Qtborder}
Q^\tb \sim \frac{qR}{L_{T_i}} \left ( \frac{\rho_i}{L_{T_i}} \right )^2 n_e v_{ti} T_i,
\end{equation} 
the perpendicular eddy length is
\begin{equation} \label{eq:lbotorder}
l_\bot \sim \frac{qR}{L_{T_i}} \rho_i
\end{equation}
and the size of the turbulent perturbation of the electrostatic potential $\phi^\tb$ is 
\begin{equation} \label{eq:phitborder}
\frac{e \phi^\tb}{T_i} \sim \frac{qR}{L_{T_i}} \frac{\rho_i}{L_{T_i}}.
\end{equation}
Here $n_e$ is the electron density, $T_i$ is the ion temperature, $L_{T_i} = |\nabla \ln T_i|^{-1}$ is the characteristic length of the ion temperature, and $qR$ is the tokamak connection length. Equations \eq{eq:Qtborder}, \eq{eq:lbotorder} and \eq{eq:phitborder} were obtained for turbulence in which the density and temperature fluctuations are not in phase. From equations \eq{eq:Qtborder}, \eq{eq:lbotorder} and \eq{eq:phitborder}, we find
\begin{equation}
\frac{l_\bot}{L_{T_i}} \sim \frac{e \phi^\tb}{T_i} \sim \left ( \frac{q R}{\rho_i} \right )^{1/3} \left ( \frac{Q^\tb}{n_e v_{ti} T_i} \right )^{2/3}.
\end{equation}
Global gyrokinetic codes are then useful if $Q^\tb$ is sufficiently large that $l_\bot/L_{T_i} \sim 1$ whereas at the same time, $\rho_\ast$ is still sufficiently small to justify averaging out the gyromotion. Importantly, in the limit $l_\bot/L_{T_i} \sim 1$, the turbulent fluctuations of the potential are of the order of the thermal energy of the plasma, and some terms that are neglected in $\delta f$ formulations become important. We do not attempt to model this extreme limit in this article, and for this reason, we do not consider the case $e \phi^\tb/T_i \sim 1$ any further.

Global gyrokinetic codes have proven useful for certain problems, but they have several disadvantages compared to flux tube formulations. They are computationally expensive, and unlike most flux tube codes, they do not use efficient spectral methods which are particularly useful for gyroaveraging. The time scale separation between the transport time scale and the turbulent time scale is not infinite as in flux tube formulations, and as a result, ad hoc sources and sinks of particles, momentum and energy must be included to mantain density, rotation and temperature gradients. Perhaps the greatest challenge facing global simulations is choosing suitable boundary conditions at the two flux surfaces that bound the simulation domain. The boundary conditions imposed in global $\delta f$ codes are typically chosen to be 'benign' from a numerical perspective; i.e., not chosen according to physical considerations such as regularity at the magnetic axis, or the open field lines beyond the last closed flux surface. The influence of these boundary conditions on the results of the simulations for $l_\bot \sim L$ is unclear. Boundary conditions are important in this limit, but if this is the case, the boundary conditions should be based on physical arguments.

Given these limitations, we propose a different approach to capture effects included in global $\delta f$ codes but not in flux tube formulations. Our approach exploits the advantages of the flux tube formulation, and relaxes the assumption $l_\bot/L \rightarrow 0$ by keeping more terms in the expansion in $l_\bot/L \ll 1$. 

The new method presented in this article is ideal for intrinsic rotation calculations. Intrinsic rotation is the rotation observed in tokamaks without any external momentum input \cite{rice99, rice05, bortolon06, scarabosio06, degrassie07, duval07, rice07, eriksson09, incecushman09, lin09, camenen10, solomon10, mcdermott11, rice11a, rice11b, parra12a, hillesheim14}. It is the result of the turbulent redistribution of toroidal angular momentum within the tokamak plasma. To lowest order in $\rho_\ast$, momentum redistribution is not possible in up-down symmetric tokamaks due to a symmetry of the gyrokinetic equations \cite{peeters05, parra11c, sugama11a}. Intrinsic rotation can only be driven by mechanisms that break this symmetry, such as up-down asymmetry of the magnetic flux surfaces \cite{camenen09b, camenen09c, ball14}, or higher order terms in $\rho_\ast$ \cite{parra10b, parra12b}. Some of the higher order terms that have been considered in previous work are: RF heating and current drive \cite{perkins01, lee12}, the radial variation of the gradients of density and temperature \cite{diamond08, gurcan10, waltz11, camenen11}, the neoclassical flows of particles and heat parallel to the flux surface \cite{parra10a, parra11d, barnes13, lee14a, lee14b, lee14c}, finite orbit width effects \cite{parra10a, parra11d}, and the poloidal variation of the turbulence characteristics \cite{sung13} (see \cite{parra14b} for a rigorous treatment of all these effects for conventional tokamaks). The new method for global $\delta f$ gyrokinetics that we propose here can be used to study the effect of the radial and poloidal variation of the turbulence characteristics on rotation.

We present gyrokinetics briefly in section \ref{sec:gyrokinetics}. We then propose the new method to do global $\delta f$ gyrokinetics in section \ref{sec:newapproach}, and we prove that it is equivalent to the method implemented in current global $\delta f$ simulations for $l_\bot/L \ll 1$ in section \ref{sec:equivalence}. We finish with some remarks and conclusions in section \ref{sec:conclusions}.

\section{Gyrokinetics} \label{sec:gyrokinetics}
Gyrokinetics \cite{catto78, frieman82} averages out the gyromotion time scale while keeping finite gyroradius effects. To be able to average out the gyromotion, several terms must be small in $\rho_\ast \ll 1$. In particular, for electrostatic turbulence in a plasma with flow below the ion thermal speed, the electric field must satisfy \cite{dimits92, parra08}
\begin{equation} \label{eq:Efieldorder}
\bE = - \nabla \phi \lesssim \rho_\ast \frac{v_{ti} B}{c} \ll \frac{v_{ti} B}{c},
\end{equation}
where $B$ is the magnitude of the magnetic field, and $c$ is the speed of light. When this bound is satisfied by the electric field, the modifications to the lowest order circular gyro orbit are small in $\rho_\ast$, and the calculation of the particle orbit can be done order by order in $\rho_\ast$ to the desired accuracy. The methods to perform the expansion are varied \cite{XSLee83, WWLee83, dubin83, hahm88, sugama96, sugama97, sugama98, brizard07, parra08, abel13, parra11a}. Once the particle motion is known, it is straightforward to split it into secular drifts and periodic oscillations, and it is possible to average over the periodic oscillations because they happen in the fast gyromotion time scale. For a recent review on the topic of gyrokinetics, see \cite{krommes12}.

Because we want to keep finite gyroradius corrections, we allow wavelengths perpendicular to the magnetic field of the order of $\rho_i$, that is, $\rho_i/l_\bot \sim 1$. According to the bound \eq{eq:Efieldorder}, the components of $\phi$ that have such short wavelengths are of order $e \phi/T_e \lesssim \rho_\ast$. The turbulence wavelength parallel to the magnetic field is much larger than the perpendicular wavelength, $l_{||}/l_\bot \sim \rho_\ast^{-1} \gg 1$. The reason for this difference between the parallel and the perpendicular scale lengths is the disparate size of the parallel and perpendicular velocities of the secular particle motion. In the parallel direction, the particle moves with its full parallel velocity, whereas in the perpendicular direction the particle velocity almost averages to zero, giving perpendicular drifts slower than the thermal speed by a factor of $\rho_\ast \ll 1$.

To describe the particle motion, new phase space coordinates are defined order by order in $\rho_\ast$ \cite{parra08}. We use the guiding center position $\bR$, the parallel velocity $v_{||}$, the magnetic moment $\mu$ and the gyrophase $\varphi$. These variables are defined such that the particle's position and velocity are
\begin{equation} \label{eq:rdef}
\boldr \simeq \bR + \rhobf (\bR, \mu, \varphi)
\end{equation}
and
\begin{equation} \label{eq:vdef}
\bv \simeq v_{||} \bun (\bR) + \frac{Z_s e}{m_s c} \rhobf(\bR, \mu, \varphi) \times \bB(\bR)
\end{equation}
to lowest order in $\rho_\ast$. Here \begin{equation}
\rhobf (\bR, \mu, \varphi) = \frac{m_s c}{Z_s e} \sqrt{\frac{2\mu}{B(\bR)}} ( - \sin \varphi\, \eun_1(\bR) + \cos \varphi\, \eun_2 (\bR) )
\end{equation}
is the gyroradius, $\bun = \bB/B$ is the unit vector in the direction of the magnetic field $\bB$, and $\eun_1$ and $\eun_2$ are two unit vector that form an orthonormal set with $\bun$ and satisfy $\eun_1 \times \eun_2 = \bun$. Of these coordinates, only the gyrophase $\varphi$ changes in the fast gyromotion time scale. The other coordinates are defined such that they do not vary in the gyromotion time scale. As a result, we can average out the gyromotion by just averaging over $\varphi$. The distribution function will not depend on the gyrophase to lowest order because of this averaging procedure. Expressions \eq{eq:rdef} and \eq{eq:vdef} are only valid to lowest order in $\rho_\ast$. For a complete and consistent treatment of the gyrokinetic transformation to the correct order, see \cite{parra11a}, where the transformation is calculated to second order in $\rho_\ast$.

Before writing the gyrokinetic system of equations, we make an additional assumption, namely the turbulent fluctuations are small compared to the background. This approximation is valid in the core of tokamaks and stellarators \cite{mckee01}. In addition, in the core, the collision frequency is much larger than the inverse of the transport time scale, making the background distribution function a Maxwellian to lowest order. Then, the distribution function for species $s$ is
\begin{equation} \label{eq:fssplit}
f_s (\bR, v_{||}, \mu, t) = f_{Ms} (\bR, v_{||}, \mu) + f_s^\tb (\bR, v_{||}, \mu, t),
\end{equation}
where
\begin{equation}
\fl f_{Ms} (\bR, v_{||}, \mu) = n_s ( \psi (\bR) ) \left ( \frac{m_s}{2\pi T_s (\psi(\bR))} \right )^{3/2} \exp \left ( - \frac{m_s (v_{||}^2 + 2\mu B(\bR))}{2T_s(\psi(\bR))} \right )
\end{equation}
is a stationary Maxwellian and $f_s^\tb (\bR, v_{||}, \mu, t)$ is the turbulent piece of the distribution function. The density $n_s(\psi)$ and the temperature $T_s(\psi)$ are flux functions, that is, they only depend on our radial coordinate $\psi$, the poloidal magnetic flux divided by $2\pi$. Correspondingly, the electrostatic potential is 
\begin{equation} \label{eq:phisplit}
\phi (\boldr, t) = \phi^\lw (\psi(\boldr) ) + \phi^\tb (\boldr, t),
\end{equation}
where the background potentical $\phi^\lw (\psi)$ is a flux function because of quasineutrality, and $\phi^\tb (\boldr, t)$ is the turbulent piece of the potential. Both $\phi^\lw$ and $f_{Ms}$ are slowly varying in space. In particular, $f_{Ms}$ can be Taylor expanded around $\boldr$ to give
\begin{eqnarray}
f_{Ms} \simeq n_s ( \psi (\boldr) ) \left ( \frac{m_s}{2\pi T_s (\psi(\boldr))} \right )^{3/2} \exp \left ( - \frac{m_s v^2}{2T_s(\psi(\boldr))} \right ).
\end{eqnarray}

Using expressions \eq{eq:fssplit} and \eq{eq:phisplit}, we obtain the lowest order Fokker-Planck equation for $f_s^\tb$,
\begin{eqnarray} \label{eq:FPequation}
\fl \frac{\partial h_s^\tb}{\partial t} + \left ( v_{||} \bun + \bv_{Ms} - \frac{c}{B} \nabla_\bR \phi^\lw \times \bun - \frac{c}{B} \nabla_\bR \langle \phi^\tb \rangle \times \bun \right ) \cdot \nabla_\bR h_s^\tb \nonumber\\ - \mu \bun \cdot \nabla_\bR B \frac{\partial h_s^\tb}{\partial v_{||}} - \frac{c}{B} (\nabla_\bR \langle \phi^\tb \rangle \times \bun) \cdot \nabla_\bR \psi \Bigg [ \frac{1}{n_s} \frac{\partial n_s}{\partial \psi} + \frac{Z_s e}{T_s} \frac{\partial \phi^\lw}{\partial \psi} \nonumber\\ + \left ( \frac{m_s (v_{||}^2 + 2\mu B)}{2T_s} - \frac{3}{2} \right ) \frac{1}{T_s} \frac{\partial T_s}{\partial \psi} \Bigg ] f_{Ms} - \frac{Z_s e}{T_s} \frac{\partial \langle \phi^\tb \rangle}{\partial t} f_{Ms} = 0,
\end{eqnarray}
where we have neglected collisions for simplicity, and we have written the equation in terms of the non-adiabatic response
\begin{equation}
h_s^\tb = f_s^\tb + \frac{Z_s e \langle \phi^\tb \rangle}{T_s} f_{Ms},
\end{equation}
and the lowest order quasineutrality equation for $\phi^\tb$,
\begin{eqnarray} \label{eq:QNequation}
\fl \sum_s Z_s \int B\, h_s^\tb ( \boldr - \rhobf(\boldr, \mu, \varphi), v_{||}, \mu, t)\, \dd v_{||}\, \dd \mu\, \dd \varphi - \sum_s \frac{Z_s^2 n_s e \phi^\tb}{T_s} = 0.
\end{eqnarray}
Here
\begin{equation}
\bv_{Ms} = \frac{v_{||}^2}{\Omega_s} \bun \times \kappabf + \frac{\mu}{\Omega_s} \bun \times \nabla_\bR B
\end{equation}
is the curvature and magnetic drift, $\kappabf = \bun \cdot \nabla_\bR \bun$ is the curvature of the magnetic field line, and $\langle \ldots \rangle$ is the gyrophase average holding $\bR$, $v_{||}$, $\mu$ and $t$ fixed. In particular, the gyroaveraged potential is
\begin{equation}
\langle \phi^\tb \rangle (\bR, \mu, t) = \frac{1}{2\pi} \int_0^{2\pi} \phi^\tb ( \bR + \rhobf(\bR, \mu, \varphi), t )\, \dd \varphi.
\end{equation}
Equations \eq{eq:FPequation} and \eq{eq:QNequation} are only correct to lowest order in $\rho_\ast$, but they are the most commonly used equations in global $\delta f$ codes. The higher order terms that we keep in the next sections are the ones needed to capture the physics included in global $\delta f$ codes. For other higher order effects, see \cite{parra11a, parra14b}. 

\section{New approach to global $\delta f$ gyrokinetics} \label{sec:newapproach}
Because we need to distinguish between the directions perpendicular and parallel to the magnetic field due to the disparate length scales, we use the set of flux coordinates $\{ \psi, \alpha, \theta \}$ to describe the spatial dependence of $h_s^\tb$ and $\phi^\tb$. As explained above, $\psi$ is the poloidal magnetic flux divided by $2\pi$, and determines the flux surface. The angle $\alpha$ is the Clebsch variable corresponding to $\psi$, 
\begin{equation}
\bB = \nabla \alpha \times \nabla \psi,
\end{equation}
and identifies a magnetic field line within the flux surface. The two coordinates $\psi$ and $\alpha$ describe the direction perpendicular to the magnetic field and satisfy $\bB \cdot \nabla \psi = 0 = \bB \cdot \nabla \alpha$. We need one coordinate to determine the position along the magnetic field line. We choose the poloidal angle $\theta$.

With periodic boundary conditions, it is possible to write the perpendicular spatial dependence of the potential and the distribution function as an eikonal, that is,
\begin{equation} \label{eq:phitbscales}
\fl \phi^\tb (\boldr, t) = \sum_{k_\psi, k_\alpha} \underline{\phi}^\tb (k_\psi, k_\alpha, \psi(\boldr), \alpha(\boldr), \theta(\boldr), t) \exp(i k_\psi \psi(\boldr) + i k_\alpha \alpha (\boldr) )
\end{equation}
and
\begin{eqnarray} \label{eq:fstbscales}
\fl h_s^\tb (\bR, v_{||}, \mu, t) = \sum_{k_\psi, k_\alpha} \underline{h}_s^\tb (k_\psi, k_\alpha, \psi(\bR), \alpha(\bR), \theta(\bR), v_{||}, \mu, t) \nonumber\\ \times \exp(i k_\psi \psi(\bR) + i k_\alpha \alpha (\bR) ).
\end{eqnarray}
The wavenumbers $k_\psi$ and $k_\alpha$ are ordered such that
\begin{equation} \label{eq:kbotdef}
\bk_\bot (\boldr) = k_\psi \nabla \psi + k_\alpha \nabla \alpha
\end{equation}
is of order $\rho_i^{-1}$, that is, the eikonal $k_\psi \psi + k_\alpha \alpha$ gives the fast spatial dependence of the turbulent fluctuations. In addition to the eikonal, in the Fourier coefficients $\underline{\phi}^\tb$ and $\underline{h}_s^\tb$ in \eq{eq:phitbscales} and \eq{eq:fstbscales}, we have kept a slow dependence on the position. The characteristic length of this dependence is of the order of the size of the machine,
\begin{equation}
\nabla \ln \underline{\phi}^\tb \sim \nabla_\bR \ln \underline{h}_s^\tb \sim \frac{1}{L},
\end{equation}
where
\begin{equation}
\nabla = \nabla \psi \frac{\partial}{\partial \psi} + \nabla \alpha \frac{\partial}{\partial \alpha} + \nabla \theta \frac{\partial}{\partial \theta},
\end{equation}
and the equivalent formula is used for $\nabla_\bR$. This slow spatial dependence represents two different phenomena: the long parallel wavelengths of the turbulence, and the slow dependence of the turbulence characteristics on the spatial location. The long parallel wavelengths of the turbulence are included in flux tube codes, but the effect of the slow variation of the turbulence characteristics across the machine is not included. Note that in tokamaks, $\partial \underline{\phi}^\tb/\partial \alpha = 0 = \partial \underline{h}^\tb_s/\partial \alpha$ due to axisymmetry.

Using the forms in \eq{eq:phitbscales} and \eq{eq:fstbscales}, we obtain
\begin{equation} \label{eq:phiaveapprox}
\langle \phi^\tb \rangle \simeq \sum_{k_\psi, k_\alpha} ( \underline{\phi}^\tb J_0 (\Lambda_s) + \underline{\Delta \langle \phi^\tb \rangle} ) \exp ( i k_\psi \psi (\bR) + i k_\alpha \alpha (\bR))
\end{equation}
and
\begin{eqnarray} \label{eq:hrvapprox}
\fl \int_0^{2\pi} h_s^\tb ( \boldr - \rhobf (\boldr, \mu, \varphi), v_{||}, \mu, t) \, \dd \varphi \simeq 2\pi \sum_{k_\psi, k_\alpha} ( \underline{h}_s^\tb J_0 (\lambda_s) + \underline{\Delta h}_s^\tb ) \nonumber \\ \times \exp ( i k_\psi \psi (\boldr) + i k_\alpha \alpha (\boldr)),
\end{eqnarray}
where $J_n$ is the $n$-th order Bessel function of the first kind,
\begin{equation}
\Lambda_s (\bR, \mu) = \frac{m_s c k_\bot (\bR)}{Z_s e} \sqrt{\frac{2\mu}{B(\bR)}},
\end{equation}
\begin{equation}
\lambda_s (\boldr, \mu) = \Lambda_s(\boldr, \mu) = \frac{k_\bot (\boldr) v_\bot}{\Omega_s (\boldr)}
\end{equation}
and $k_\bot$ is the magnitude of $\bk_\bot$ in \eq{eq:kbotdef}. The corrections $\underline{\Delta h}_s^\tb$ and $\underline{\Delta \langle \phi^\tb \rangle}$ are small in $\rho_\ast$, and we have neglected terms that are of order $\rho_\ast^2$ or higher. The corrections $\underline{\Delta h}_s^\tb$ and $\underline{\Delta \langle \phi^\tb \rangle}$ are given by
\begin{eqnarray}
\fl \underline{\Delta \langle \phi^\tb \rangle} = \frac{2J_1(\Lambda_s)}{\Lambda_s}\frac{i \mu B}{2 \Omega_s^2} \left ( \matI - \bun \bun \right ) : \nabla_\bR \bk_\bot\, \underline{\phi}^\tb - G(\Lambda_s) \frac{i k_\bot \mu^2 B^2 }{4\Omega_s^4 }  \bk_\bot \cdot \nabla_\bR k_\bot\, \underline{\phi}^\tb  \nonumber\\ + \frac{2 J_1 (\Lambda_s)}{\Lambda_s} \frac{i \mu B}{\Omega_s^2}  \bk_\bot \cdot \nabla_\bR \underline{\phi}^\tb
\end{eqnarray}
and
\begin{eqnarray}
\fl \underline{\Delta h}_s^\tb =  \frac{2J_1(\lambda_s)}{\lambda_s}\frac{i \mu B}{2 \Omega_s^2} \left ( \matI - \bun \bun \right ) : \nabla_\bR \bk_\bot\, \underline{h}_s^\tb - G(\lambda_s) \frac{i k_\bot \mu^2 B^2 }{4\Omega_s^4 }  \bk_\bot \cdot \nabla_\bR k_\bot\, \underline{h}_s^\tb  \nonumber\\ + \frac{2 J_1 (\lambda_s)}{\lambda_s} \frac{i \mu B}{\Omega_s^2}  \bk_\bot \cdot \nabla_\bR \underline{h}_s^\tb,
\end{eqnarray}
where $\matI$ is the unit matrix, our convention for double contraction is $\mathbf{b} \mathbf{a} : \matrixtop{\mathbf{M}} = \mathbf{a} \cdot \matrixtop{\mathbf{M}} \cdot \mathbf{b}$, and
\begin{equation}
G(x) = \frac{8 J_1 (x) - 4x (J_0 (x) - J_2 (x))}{x^3}
\end{equation}
is defined such that $G \rightarrow 1$ for $x \rightarrow 0$. 

With the results in \eq{eq:phiaveapprox} and \eq{eq:hrvapprox}, and the decomposition in \eq{eq:phitbscales} and \eq{eq:fstbscales}, equations \eq{eq:FPequation} and \eq{eq:QNequation} become
\begin{eqnarray} \label{eq:FPfluxtube}
\fl \frac{\partial \underline{h}_s^\tb}{\partial t} + \left ( v_{||} \bun \cdot \nabla_\bR \theta \frac{\partial}{\partial \theta} - \mu \bun \cdot \nabla_\bR B \frac{\partial}{\partial v_{||}} \right ) \underline{h}_s^\tb + i \left ( - k_\alpha c \frac{\partial \phi^\lw}{\partial \psi} + \bk_\bot \cdot \bv_{Ms} \right ) \underline{h}_s^\tb \nonumber\\ + c \sum_{k_\psi^\prime, k_\alpha^\prime} (k_\psi^\prime k_\alpha^{\prime\prime} - k_\alpha^\prime k_\psi^{\prime\prime}) J_0 (\Lambda_s^\prime) (\underline{\phi}^\tb)^\prime (\underline{h}_s^\tb)^{\prime\prime} - \frac{Z_s e}{T_s} J_0(\Lambda_s) f_{Ms} \frac{\partial \underline{\phi}^\tb}{\partial t} \nonumber\\ + i k_\alpha c \Bigg ( \frac{1}{n_s} \frac{\partial n_s}{\partial \psi} + \frac{Z_s e}{T_s} \frac{\partial \phi^\lw}{\partial \psi} + \left ( \frac{m_s (v_{||}^2 + 2\mu B)}{2T_s} - \frac{3}{2} \right ) \frac{1}{T_s} \frac{\partial T_s}{\partial \psi} \Bigg ) J_0(\Lambda_s) f_{Ms} \underline{\phi}^\tb \nonumber\\ =  - \left ( - \frac{c}{B} \nabla_\bR \phi^\lw \times \bun + \bv_{Ms} \right ) \cdot \nabla_\bR \underline{h}_s^\tb  + \frac{Z_s e}{T_s}  f_{Ms} \frac{\partial \underline{\Delta \langle \phi^\tb \rangle}}{\partial t} \nonumber\\ - i k_\alpha c \Bigg ( \frac{1}{n_s} \frac{\partial n_s}{\partial \psi} + \frac{Z_s e}{T_s} \frac{\partial \phi^\lw}{\partial \psi} + \left ( \frac{m_s (v_{||}^2 + 2\mu B)}{2T_s} - \frac{3}{2} \right ) \frac{1}{T_s} \frac{\partial T_s}{\partial \psi} \Bigg ) f_{Ms} \underline{\Delta \langle \phi^\tb \rangle} \nonumber\\ - \Bigg ( \frac{1}{n_s} \frac{\partial n_s}{\partial \psi} + \frac{Z_s e}{T_s} \frac{\partial \phi^\lw}{\partial \psi} + \left ( \frac{m_s (v_{||}^2 + 2\mu B)}{2T_s} - \frac{3}{2} \right ) \frac{1}{T_s} \frac{\partial T_s}{\partial \psi} \Bigg ) \nonumber\\ \times \frac{c}{B} (\nabla_\bR \psi \times \bun) \cdot \nabla_\bR ( \underline{\phi}^\tb J_0 (\Lambda_s) ) f_{Ms} \nonumber\\ + \sum_{k_\psi^\prime, k_\alpha^\prime} \Bigg [- c (k_\psi^\prime k_\alpha^{\prime\prime} - k_\alpha^\prime k_\psi^{\prime\prime}) (\underline{\Delta \langle \phi^\tb \rangle})^\prime (\underline{h}_s^\tb)^{\prime\prime} \nonumber\\ + \frac{i c}{B} (\underline{\phi}_1^\tb)^\prime J_0 ( \Lambda_s^\prime ) (\bk^\prime_\bot \times \bun) \cdot \nabla_\bR (\underline{f}_{s1}^\tb)^{\prime\prime} \nonumber\\ - \frac{i c}{B}  (\underline{f}_{s1}^\tb)^\prime(\bk_\bot^\prime \times \bun) \cdot \nabla_\bR ( (\underline{\phi}_1^\tb)^{\prime\prime} J_0 ( \Lambda_s^{\prime\prime} ) ) \Bigg ]
\end{eqnarray}
and
\begin{eqnarray} \label{eq:QNfluxtube}
\fl 2\pi \sum_s Z_s \int B \, \underline{h}_s^\tb J_0 (\lambda_s)\, \dd v_{||}\, \dd \mu - \sum_s \frac{Z_s^2 n_s e \underline{\phi}^\tb}{T_s} = - 2 \pi \sum_s Z_s \int B \, \underline{\Delta h}_s^\tb\, \dd v_{||}\, \dd \mu.
\end{eqnarray}
It is important to remember that all the coefficients in these equations depend on $\psi$ and $\alpha$ even though that dependence is not used in flux tube formulations. A prime on a function such as $\underline{\phi}^\tb$ or $\Lambda_s$ indicates that it depends on $k_\psi^\prime$ and $k_\alpha^\prime$, e.g., $(\underline{\phi}^\tb)^\prime = \underline{\phi}^\tb (k_\psi^\prime, k_\alpha^\prime, \psi(\bR), \alpha(\bR), \theta(\bR), t)$. Two primes indicate that it depends on $k_\psi^{\prime\prime}$ and $k_\alpha^{\prime\prime}$, e.g., $(\underline{\phi}^\tb)^{\prime\prime} = \underline{\phi}^\tb (k_\psi^{\prime\prime}, k_\alpha^{\prime\prime}, \psi(\bR), \alpha(\bR), \theta(\bR), t)$, where
\begin{equation}
k_\psi^{\prime\prime} = k_\psi - k_\psi^\prime,
\end{equation}
and
\begin{equation}
k_\alpha^{\prime\prime} = k_\alpha - k_\alpha^\prime.
\end{equation} 
In \eq{eq:FPfluxtube} and \eq{eq:QNfluxtube}, the terms that are small because they correspond to the slow derivatives are on the right side of the equations. In flux tube simulations, these terms are neglected, leaving the simpler equations
\begin{eqnarray} \label{eq:FPfluxtube0}
\fl \frac{\partial \underline{h}_s^\tb}{\partial t} + \left ( v_{||} \bun \cdot \nabla_\bR \theta \frac{\partial}{\partial \theta} - \mu \bun \cdot \nabla_\bR B \frac{\partial}{\partial v_{||}} \right ) \underline{h}_s^\tb + i \left ( - k_\alpha c \frac{\partial \phi_0}{\partial \psi} + \bk_\bot \cdot \bv_{Ms} \right ) \underline{h}_s^\tb \nonumber\\ + c \sum_{k_\psi^\prime, k_\alpha^\prime} (k_\psi^\prime k_\alpha^{\prime\prime} - k_\alpha^\prime k_\psi^{\prime\prime})  J_0 (\Lambda_s^\prime) (\underline{\phi}^\tb)^\prime (\underline{h}_s^\tb)^{\prime\prime} - \frac{Z_s e}{T_s}  J_0(\Lambda_s) f_{Ms} \frac{\partial \underline{\phi}^\tb}{\partial t} \nonumber\\ + i k_\alpha c \Bigg ( \frac{1}{n_s} \frac{\partial n_s}{\partial \psi} + \frac{Z_s e}{T_s} \frac{\partial \phi^\lw}{\partial \psi} + \left ( \frac{m_s (v_{||}^2 + 2\mu B)}{2T_s} - \frac{3}{2} \right ) \frac{1}{T_s} \frac{\partial T_s}{\partial \psi} \Bigg ) \nonumber\\ \times  J_0(\Lambda_s) f_{Ms} \underline{\phi}^\tb \simeq 0
\end{eqnarray}
and
\begin{eqnarray} \label{eq:QNfluxtube0}
2\pi \sum_s Z_s \int B \, \underline{h}_s^\tb J_0 (\lambda_s)\, \dd v_{||}\, \dd \mu - \sum_s \frac{Z_s^2 n_s e \underline{\phi}^\tb}{T_s} \simeq 0.
\end{eqnarray}
If the terms on the right side of equations \eq{eq:FPfluxtube} and \eq{eq:QNfluxtube} are implemented into a flux tube code, they give the next order corrections in the expansion in $l_\bot/L \ll 1$, and some of the terms in the expansion in $\rho_\ast$, but not all of them. To calculate the right side of equations \eq{eq:FPfluxtube} and \eq{eq:QNfluxtube}, we need the slow derivatives $\partial \underline{h}_s^\tb/\partial \psi$, $\partial \underline{h}_s^\tb/\partial \alpha$, $\partial \underline{\phi}^\tb/\partial \psi$ and $\partial \underline{\phi}^\tb/\partial \alpha$. The slow derivatives $\partial \underline{h}_s^\tb/\partial \theta$ and $\partial \underline{\phi}^\tb/\partial \theta$ are already determined by the lowest order equations \eq{eq:FPfluxtube0} and \eq{eq:QNfluxtube0}. To determine $\partial \underline{h}_s^\tb/\partial \psi$, $\partial \underline{h}_s^\tb/\partial \alpha$, $\partial \underline{\phi}^\tb/\partial \psi$ and $\partial \underline{\phi}^\tb/\partial \alpha$, one can run several flux tube simulations to obtain $\underline{h}_s^\tb ( k_\psi, k_\alpha, \psi(\bR), \alpha(\bR), \theta(\bR), v_{||}, \mu, t)$ and $\underline{\phi}^\tb ( k_\psi, k_\alpha, \psi(\boldr), \alpha(\boldr), \theta(\boldr), t)$ at different spatial locations, and then differentiate numerically. Another possibility is to integrate in time $\partial \underline{h}_s^\tb/\partial \psi$, $\partial \underline{h}_s^\tb/\partial \alpha$, $\partial \underline{\phi}^\tb/\partial \psi$ and $\partial \underline{\phi}^\tb/\partial \alpha$ using the derivatives of the lowest order equations \eq{eq:FPfluxtube0} and \eq{eq:QNfluxtube0}. The time evolution equations for $\partial \underline{h}_s^\tb/\partial \psi$, $\partial \underline{h}_s^\tb/\partial \alpha$, $\partial \underline{\phi}^\tb/\partial \psi$ and $\partial \underline{\phi}^\tb/\partial \alpha$ are long and for that reason are given in \ref{app:ddpsiddaphaequations}, in equations \eq{eq:FPgradpsi}-\eq{eq:QNgradalpha}.

We have used the lowest order equations \eq{eq:FPfluxtube0} and \eq{eq:QNfluxtube0} to derive equations for $\partial \underline{h}_s^\tb/\partial \psi$, $\partial \underline{h}_s^\tb/\partial \alpha$, $\partial \underline{\phi}^\tb/\partial \psi$ and $\partial \underline{\phi}^\tb/\partial \alpha$ because these gradients are not needed to higher order. They only appear in the higher order terms of \eq{eq:FPfluxtube} and \eq{eq:QNfluxtube}. If we take the spatial derivatives of \eq{eq:FPfluxtube} and \eq{eq:QNfluxtube} instead of the derivatives of the lowest order equations \eq{eq:FPfluxtube0} and \eq{eq:QNfluxtube0}, $\partial \underline{h}_s^\tb/\partial \psi$, $\partial \underline{h}_s^\tb/\partial \alpha$, $\partial \underline{\phi}^\tb/\partial \psi$ and $\partial \underline{\phi}^\tb/\partial \alpha$ can be known to an order higher in $l_\bot/L \ll 1$ and as a result, $\underline{h}_s^\tb$ and $\underline{\phi}^\tb$ can be determined to $O(l_\bot^2/L^2)$. In this case, the equations for $\partial \underline{h}_s^\tb/\partial \psi$, $\partial \underline{h}_s^\tb/\partial \alpha$, $\partial \underline{\phi}^\tb/\partial \psi$ and $\partial \underline{\phi}^\tb/\partial \alpha$ include terms with second spatial derivatives of $\underline{h}^\tb_s$ and $\underline{\phi}^\tb$ with respect to $\psi$ and $\alpha$ that must be determined. To calculate these second derivatives we can use the second derivatives of the lowest order equations \eq{eq:FPfluxtube0} and \eq{eq:QNfluxtube0}, truncating the system of equations at this order, or use again the second spatial derivatives of equations \eq{eq:FPfluxtube} and \eq{eq:QNfluxtube} that in turn include the third derivatives with respect to $\psi$ and $\alpha$. The more spatial derivatives we keep of $\underline{h}_s^\tb$ and $\underline{\phi}^\tb$, the more accurate the calculation is in $l_\bot/L \ll 1$.

To summarize, to keep first order terms in $l_\bot/L \ll 1$, we must solve equations \eq{eq:FPfluxtube}, \eq{eq:QNfluxtube} and \eq{eq:FPgradpsi}-\eq{eq:QNgradalpha} simultaneously. In the next section, we show that this is equivalent to solving \eq{eq:FPequation} and \eq{eq:QNequation} with a global $\delta f$ code for $l_\bot/L \ll 1$.

\section{Equivalence between the two methods to do global $\delta f$ gyrokinetics} \label{sec:equivalence}
Instead of solving the equations given in section \ref{sec:newapproach}, one can solve equations \eq{eq:FPequation} and \eq{eq:QNequation} with a global $\delta f$ code. We already pointed out in the introduction that global $\delta f$ codes give the same solution as flux tube simulations in the limit $l_\bot/L \ll 1$ \cite{goerler11, candy04}. In this section we show that the flux tube formulation in section~\ref{sec:newapproach} gives the same result as a global $\delta f$ code to an order higher in $l_\bot/L\ll 1$ than the usual flux tube formulation in equations \eq{eq:FPfluxtube0} and \eq{eq:QNfluxtube0}.

The solutions to the lowest order flux tube equations \eq{eq:FPfluxtube0} and \eq{eq:QNfluxtube0} in a flux tube located at $\psi = \psi_0$ and $\alpha = \alpha_0$ are solutions to the global $\delta f$ equations \eq{eq:FPequation} and \eq{eq:QNequation} to lowest order in $l_\bot/L$ in a region of size $l_\bot$ around $\psi = \psi_0$ and $\alpha = \alpha_0$. To show that the flux tube formulation in section \ref{sec:newapproach} gives the solution to higher order in $l_\bot/L$, we construct solutions $h_s^\tb$ and $\phi^\tb$ to the global $\delta f$ equations \eq{eq:FPequation} and \eq{eq:QNequation} from the functions $\underline{h}_s^\tb$, $\underline{\phi}^\tb$, $\partial \underline{h}_s^\tb/\partial \psi$, $\partial \underline{h}_s^\tb/\partial \alpha$, $\partial \underline{\phi}^\tb/\partial \psi$ and $\partial \underline{\phi}^\tb/\partial \alpha$ obtained using the flux tube equations \eq{eq:FPfluxtube}, \eq{eq:QNfluxtube} and \eq{eq:FPgradpsi}-\eq{eq:QNgradalpha}. These solutions are
\begin{eqnarray} \label{eq:hhatdef}
\fl h_s^\tb (\bR, v_{||}, \mu, t) = \hat{h}_s^\tb  (\bR, v_{||}, \mu, t) + (\psi (\bR) - \psi_0) H_{s,\psi}^\tb (\bR, v_{||}, \mu, t) \nonumber\\ + (\alpha (\bR) - \alpha_0) H_{s,\alpha}^\tb (\bR, v_{||}, \mu, t)
\end{eqnarray}
and
\begin{equation} \label{eq:phihatdef}
\phi^\tb (\boldr, t) = \hat{\phi}^\tb (\boldr, t) + (\psi (\boldr) - \psi_0) \Phi_{s,\psi}^\tb (\boldr, t) + (\alpha (\boldr) - \alpha_0) \Phi_{s,\alpha}^\tb (\boldr, t).
\end{equation}
The functions $\hat{h}_s^\tb$ and $\hat{\phi}^\tb$ are constructed from the Fourier coefficients that are the solution to the higher order flux tube equations \eq{eq:FPfluxtube} and \eq{eq:QNfluxtube} in the flux tube around $\psi = \psi_0$ and $\alpha = \alpha_0$,
\begin{equation} \label{eq:Fourierhhat}
\fl \hat{h}_s^\tb (\bR, v_{||}, \mu, t) = \sum_{k_\psi, k_\alpha} \underline{h}_s^\tb (k_\psi, k_\alpha, \theta(\bR), v_{||}, \mu, t) \exp (i k_\psi \psi(\bR) + i k_\alpha \alpha(\bR))
\end{equation}
and
\begin{equation} \label{eq:Fourierphihat}
\hat{\phi}^\tb (\boldr, t) = \sum_{k_\psi, k_\alpha} \underline{\phi}^\tb (k_\psi, k_\alpha, \theta(\boldr), t) \exp (i k_\psi \psi(\boldr) + i k_\alpha \alpha(\boldr)).
\end{equation}
The functions $H_{s, \psi}^\tb$, $\Phi_\psi^\tb$, $H_{s, \alpha}^\tb$ and $\Phi_\alpha^\tb$ are constructed from the Fourier coefficients that are the solution to the flux tube equations \eq{eq:FPgradpsi}-\eq{eq:QNgradalpha} in the same flux tube,
\begin{equation} \label{eq:FourierHpsi}
\fl H_{s,\psi}^\tb (\bR, v_{||}, \mu, t) = \sum_{k_\psi, k_\alpha} \frac{\partial \underline{h}_s^\tb}{\partial \psi} (k_\psi, k_\alpha, \theta(\bR), v_{||}, \mu, t) \exp (i k_\psi \psi(\bR) + i k_\alpha \alpha(\bR)),
\end{equation}
\begin{equation} \label{eq:FourierPhipsi}
\Phi_\psi^\tb (\boldr, t) = \sum_{k_\psi, k_\alpha} \frac{\partial \underline{\phi}^\tb}{\partial \psi} (k_\psi, k_\alpha, \theta(\boldr), t) \exp (i k_\psi \psi(\boldr) + i k_\alpha \alpha(\boldr)),
\end{equation}
\begin{equation} \label{eq:FourierHalpha}
\fl H_{s, \alpha}^\tb (\bR, v_{||}, \mu, t) = \sum_{k_\psi, k_\alpha} \frac{\partial \underline{h}_s^\tb}{\partial \alpha} (k_\psi, k_\alpha, \theta(\bR), v_{||}, \mu, t) \exp (i k_\psi \psi(\bR) + i k_\alpha \alpha(\bR))
\end{equation}
and
\begin{equation} \label{eq:FourierPhialpha}
\Phi_\alpha^\tb (\boldr, t) = \sum_{k_\psi, k_\alpha} \frac{\partial \underline{\phi}^\tb}{\partial \alpha} (k_\psi, k_\alpha, \theta(\boldr), t) \exp (i k_\psi \psi(\boldr) + i k_\alpha \alpha(\boldr)).
\end{equation}
The functions $H_{s, \psi}^\tb$, $\Phi_\psi^\tb$, $H_{s, \alpha}^\tb$ and $\Phi_\alpha^\tb$ correspond to the slow spatial derivatives used in section \ref{sec:newapproach}. We have not used the usual derivative notation, i.e., $\partial h_s^\tb/\partial \psi$, $\partial \phi^\tb/\partial \psi$, $\partial h_s^\tb/\partial \alpha$ and $\partial \phi^\tb/\partial \alpha$, because in global $\delta f$ formulations there is no clear distinction between fast and slow spatial derivatives. In fact, the separation between fast and slow spatial dependence is somewhat arbitrary. This freedom is apparent in our formulation because the functions $H_{s, \psi}^\tb$, $\Phi_\psi^\tb$, $H_{s, \alpha}^\tb$ and $\Phi_\alpha^\tb$ are only defined by the time evolution equations they satisfy (the Fourier transforms of equations \eq{eq:FPgradpsi}-\eq{eq:QNgradalpha}). The infinite choices for the initial condition for $H_{s, \psi}^\tb$, $\Phi_\psi^\tb$, $H_{s, \alpha}^\tb$ and $\Phi_\alpha^\tb$ are a reflection of the fact that the separation between fast and slow spatial dependences is arbitrary. Conversely, the time evolution equations for $H_{s, \psi}^\tb$, $\Phi_\psi^\tb$, $H_{s, \alpha}^\tb$ and $\Phi_\alpha^\tb$ are unique.

The details of the proof that the functions in \eq{eq:hhatdef} and \eq{eq:phihatdef} are solutions to the global $\delta f$ equations \eq{eq:FPequation} and \eq{eq:QNequation} accurate to first order in $l_\bot/L$ in a region of size $l_\bot$ around $\psi = \psi_0$ and $\alpha = \alpha_0$ are given in \ref{app:proof}. The proof consists of substituting the functions $h_s^\tb$ and $\phi^\tb$ in \eq{eq:hhatdef} and \eq{eq:phihatdef} into the global $\delta f$ equations \eq{eq:FPequation} and \eq{eq:QNequation}. Using the time evolution equations for $H_{s, \psi}^\tb$, $\Phi_\psi^\tb$, $H_{s, \alpha}^\tb$ and $\Phi_\alpha^\tb$ (the Fourier transforms of equations \eq{eq:FPgradpsi}-\eq{eq:QNgradalpha}), it is possible to show that equations \eq{eq:FPequation} and \eq{eq:QNequation} are satisfied to first order in $l_\bot/L$ in a region of size $l_\bot$ around $\psi = \psi_0$ and $\alpha = \alpha_0$. It is crucial for the proof that the geometrical coefficients in the global $\delta f$ equations \eq{eq:FPequation} and \eq{eq:QNequation} can be Taylor expanded around $\psi = \psi_0$ and $\alpha = \alpha_0$. Therefore, equations \eq{eq:FPfluxtube}, \eq{eq:QNfluxtube} and \eq{eq:FPgradpsi}-\eq{eq:QNgradalpha} are equivalent to the gyrokinetic equations in a global $\delta f$ simulation as long as the coefficients in the equation are sufficiently regular that they can be Taylor expanded in $\psi$ and $\alpha$.

\section{Conclusions} \label{sec:conclusions}

We have derived a system of equations, formed by equations \eq{eq:FPfluxtube}, \eq{eq:QNfluxtube} and \eq{eq:FPgradpsi}-\eq{eq:QNgradalpha}, that gives the next order correction in $l_\bot/L \ll 1$ to the flux tube gyrokinetic equations. Equations \eq{eq:FPfluxtube}, \eq{eq:QNfluxtube} and \eq{eq:FPgradpsi}-\eq{eq:QNgradalpha} do not have to be solved in an extended radial domain. Consequently, we do not need to impose boundary conditions other than periodicity. We have shown that the system of equations \eq{eq:FPfluxtube}, \eq{eq:QNfluxtube} and \eq{eq:FPgradpsi}-\eq{eq:QNgradalpha} is equivalent to the gyrokinetic equations solved in global $\delta f$ simulations in the limit $l_\bot/L \ll 1$.

The method proposed in this article is only valid for $l_\bot/L \ll 1$, and it was developed to calculate the intrinsic rotation due to the slow spatial variation of the turbulence characteristics \cite{parra14b}. Our method cannot treat extreme cases with turbulent eddies of the order of the size of the machine because in this limit the boundary conditions at the magnetic axis and at the last closed flux surface affect the entire tokamak. Our method can be used to find the corrections due to the finite size of turbulent eddies. One possible exciting use of the system of equations \eq{eq:FPfluxtube}, \eq{eq:QNfluxtube} and \eq{eq:FPgradpsi}-\eq{eq:QNgradalpha} is to compare it with results from current global $\delta f$ codes to determine the extent to which boundary conditions affect the turbulence and break the assumptions under which we could prove that equations \eq{eq:FPfluxtube}, \eq{eq:QNfluxtube} and \eq{eq:FPgradpsi}-\eq{eq:QNgradalpha} are equivalent to a global $\delta f$ gyrokinetic simulation.

\ack{This work has been carried out within the framework of the EUROfusion Consortium and has received funding from the European Union's Horizon 2020 research and innovation programme under grant agreement number 633053. The views and opinions expressed herein do not necessarily reflect those of the European Commission. This research was supported in part by the RCUK Energy Programme (grant number EP/I501045).}

\appendix

\section{Equations for $\partial \underline{h}_s^\tb/\partial \psi$, $\partial \underline{h}_s^\tb/\partial \alpha$, $\partial \underline{\phi}^\tb/\partial \psi$ and $\partial \underline{\phi}^\tb/\partial \alpha$} \label{app:ddpsiddaphaequations}

Differentiating \eq{eq:FPfluxtube0} and \eq{eq:QNfluxtube0} with respect to $\psi$, we find
\begin{eqnarray} \label{eq:FPgradpsi}
\fl \frac{\partial}{\partial t} \left ( \frac{\partial \underline{h}_s^\tb}{\partial \psi} \right )+ \left ( v_{||} \bun \cdot \nabla_\bR \theta \frac{\partial}{\partial \theta} - \mu \bun \cdot \nabla_\bR B \frac{\partial}{\partial v_{||}} \right ) \frac{\partial \underline{h}_s^\tb}{\partial \psi} \nonumber\\ + i \left ( - k_\alpha c \frac{\partial \phi^\lw}{\partial \psi} + \bk_\bot \cdot \bv_{Ms} \right ) \frac{\partial \underline{h}_s^\tb}{\partial \psi} \nonumber\\ + c \sum_{k_\psi^\prime, k_\alpha^\prime} (k_\psi^\prime k_\alpha^{\prime\prime} - k_\alpha^\prime k_\psi^{\prime\prime})  J_0 (\Lambda_s^\prime) \Bigg [ \left (\frac{\partial \underline{\phi}^\tb}{\partial \psi} \right )^\prime(\underline{h}_s^\tb)^{\prime\prime} + (\underline{\phi}^\tb)^\prime \left (\frac{\partial \underline{h}_s^\tb}{\partial \psi} \right )^{\prime\prime} \Bigg ] \nonumber\\  - \frac{Z_s e}{T_s} J_0 (\Lambda_s) f_{Ms} \frac{\partial}{\partial t} \left (  \frac{\partial \underline{\phi}^\tb}{\partial \psi} \right ) + i k_\alpha c \Bigg ( \frac{1}{n_s} \frac{\partial n_s}{\partial \psi} + \frac{Z_s e}{T_s} \frac{\partial \phi^\lw}{\partial \psi} \nonumber\\ + \left ( \frac{m_s (v_{||}^2 + 2\mu B)}{2T_s} - \frac{3}{2} \right ) \frac{1}{T_s} \frac{\partial T_s}{\partial \psi} \Bigg ) J_0 (\Lambda_s) f_{Ms} \frac{\partial \underline{\phi}^\tb}{\partial \psi} \nonumber\\ = - v_{||} \frac{\partial}{\partial \psi} (\bun \cdot \nabla_\bR \theta) \frac{\partial\underline{h}_s^\tb}{\partial \theta} + \mu \frac{\partial}{\partial \psi} ( \bun \cdot \nabla_\bR B ) \frac{\partial \underline{h}_s^\tb}{\partial v_{||}} \nonumber\\ - \frac{\partial}{\partial \psi} \left ( -i  k_\alpha c \frac{\partial \phi^\lw}{\partial \psi} + i \bk_\bot \cdot \bv_{Ms} \right ) \underline{h}_s^\tb \nonumber\\ - c \sum_{k_\psi^\prime, k_\alpha^\prime} (k_\psi^\prime k_\alpha^{\prime\prime} - k_\alpha^\prime k_\psi^{\prime\prime}) \frac{\partial}{\partial \psi} (J_0 (\Lambda_s^\prime)) (\underline{\phi}^\tb)^\prime (\underline{h}_s^\tb)^{\prime\prime} \nonumber\\ - \frac{\partial}{\partial \psi} \Bigg [ - \frac{Z_s e}{T_s} J_0 (\Lambda_s) f_{Ms} \frac{\partial}{\partial t} + i k_\alpha c \Bigg ( \frac{1}{n_s} \frac{\partial n_s}{\partial \psi} + \frac{Z_s e}{T_s} \frac{\partial \phi^\lw}{\partial \psi} \nonumber\\ + \left ( \frac{m_s (v_{||}^2 + 2\mu B)}{2T_s} - \frac{3}{2} \right ) \frac{1}{T_s} \frac{\partial T_s}{\partial \psi} \Bigg ) J_0 (\Lambda_s) f_{Ms} \Bigg ] \underline{\phi}^\tb
\end{eqnarray}
and
\begin{eqnarray} \label{eq:QNgradpsi}
\fl 2\pi \sum_s Z_s \int B \, \frac{\partial \underline{h}_s^\tb}{\partial \psi} J_0 (\lambda_s)\, \dd v_{||}\, \dd \mu - \sum_s \frac{Z_s^2 n_s e}{T_s} \frac{\partial \underline{\phi}^\tb}{\partial \psi} = \nonumber\\ - 2\pi \sum_s Z_s \int \frac{\partial}{\partial \psi} (B  J_0 (\lambda_s)) \underline{h}_s^\tb \, \dd v_{||}\, \dd \mu + \sum_s \frac{\partial}{\partial \psi} \left (\frac{Z_s^2 n_s e}{T_s} \right ) \underline{\phi}^\tb.
\end{eqnarray}
These equations can be integrated in time to find $\partial \underline{h}_s^\tb/\partial \psi$ and $\partial \underline{\phi}^\tb/\partial \psi$. The equations for $\partial \underline{h}_s^\tb/\partial \alpha$ and $\partial \underline{\phi}^\tb/\partial \alpha$ are obtained in a similar way,
\begin{eqnarray} \label{eq:FPgradalpha}
\fl \frac{\partial}{\partial t} \left ( \frac{\partial \underline{h}_s^\tb}{\partial \alpha} \right )+ \left ( v_{||} \bun \cdot \nabla_\bR \theta \frac{\partial}{\partial \theta} - \mu \bun \cdot \nabla_\bR B \frac{\partial}{\partial v_{||}} \right ) \frac{\partial \underline{h}_s^\tb}{\partial \alpha} \nonumber\\ + i \left ( - k_\alpha c \frac{\partial \phi^\lw}{\partial \psi} + \bk_\bot \cdot \bv_{Ms} \right ) \frac{\partial \underline{h}_s^\tb}{\partial \alpha} \nonumber\\ + c \sum_{k_\psi^\prime, k_\alpha^\prime} (k_\psi^\prime k_\alpha^{\prime\prime} - k_\alpha^\prime k_\psi^{\prime\prime})  J_0 (\Lambda_s^\prime) \Bigg [ \left (\frac{\partial \underline{\phi}^\tb}{\partial \alpha} \right )^\prime (\underline{h}_s^\tb)^{\prime\prime} + (\underline{\phi}^\tb)^\prime \left (\frac{\partial \underline{\phi}^\tb}{\partial \alpha} \right )^{\prime\prime} \Bigg ] \nonumber\\ - \frac{Z_s e}{T_s} J_0 (\Lambda_s) f_{Ms} \frac{\partial}{\partial t} \left (\frac{\partial \underline{\phi}^\tb}{\partial \alpha} \right ) + i k_\alpha c \Bigg ( \frac{1}{n_s} \frac{\partial n_s}{\partial \psi} + \frac{Z_s e}{T_s} \frac{\partial \phi^\lw}{\partial \psi} \nonumber\\ + \left ( \frac{m_s (v_{||}^2 + 2\mu B)}{2T_s} - \frac{3}{2} \right ) \frac{1}{T_s} \frac{\partial T_s}{\partial \psi} \Bigg ) J_0 (\Lambda_s) f_{Ms} \frac{\partial \underline{\phi}^\tb}{\partial \alpha} \nonumber\\ = - v_{||} \frac{\partial}{\partial \alpha} (\bun \cdot \nabla_\bR \theta) \frac{\partial\underline{h}_s^\tb}{\partial \theta} + \mu \frac{\partial}{\partial \alpha} ( \bun \cdot \nabla_\bR B ) \frac{\partial \underline{h}_s^\tb}{\partial v_{||}} \nonumber\\ - \frac{\partial}{\partial \alpha} \left ( i \bk_\bot \cdot \bv_{Ms} \right ) \underline{h}_s^\tb - c \sum_{k_\psi^\prime, k_\alpha^\prime} (k_\psi^\prime k_\alpha^{\prime\prime} - k_\alpha^\prime k_\psi^{\prime\prime}) \frac{\partial}{\partial \alpha} (J_0 (\Lambda_s^\prime)) (\underline{\phi}^\tb)^\prime (\underline{h}_s^\tb)^{\prime\prime} \nonumber\\ - \frac{\partial}{\partial \alpha} \Bigg [ - \frac{Z_s e}{T_s} J_0 (\Lambda_s) f_{Ms} \frac{\partial}{\partial t} + i k_\alpha c \Bigg ( \frac{1}{n_s} \frac{\partial n_s}{\partial \psi}  + \frac{Z_s e}{T_s} \frac{\partial \phi^\lw}{\partial \psi} \nonumber\\ + \left ( \frac{m_s (v_{||}^2 + 2\mu B)}{2T_s} - \frac{3}{2} \right ) \frac{1}{T_s} \frac{\partial T_s}{\partial \psi} \Bigg ) J_0 (\Lambda_s) f_{Ms} \Bigg ] \underline{\phi}^\tb
\end{eqnarray}
and
\begin{eqnarray} \label{eq:QNgradalpha}
\fl 2\pi \sum_s Z_s \int B \, \frac{\partial \underline{h}_s^\tb}{\partial \alpha} J_0 (\lambda_s)\, \dd v_{||}\, \dd \mu - \sum_s \frac{Z_s^2 n_s e}{T_s} \frac{\partial \underline{\phi}^\tb}{\partial \alpha} = \nonumber\\ - 2\pi \sum_s Z_s \int \frac{\partial}{\partial \alpha} (B  J_0 (\lambda_s)) \underline{h}_s^\tb \, \dd v_{||}\, \dd \mu.
\end{eqnarray}
In tokamaks, $\partial \underline{h}_s^\tb/\partial \alpha = 0 =\partial \underline{\phi}^\tb/\partial \alpha$, and these last two equations are not needed.

\section{Proof that the functions defined in \eq{eq:hhatdef} and \eq{eq:phihatdef} are solutions to the global $\delta f$ equations \eq{eq:FPequation} and \eq{eq:QNequation}} \label{app:proof}

In this appendix we show that the functions $h_s^\tb$ and $\phi^\tb$ in \eq{eq:hhatdef} and \eq{eq:phihatdef} are solutions to \eq{eq:FPequation} and \eq{eq:QNequation} in a region of size $l_\bot$ around the magnetic field line $\psi = \psi_0$ and $\alpha = \alpha_0$. To describe the spatial dependence of the different coefficients in equations \eq{eq:FPequation} and \eq{eq:QNequation}, we Taylor expand these coefficients around $\psi = \psi_0$ and $\alpha = \alpha_0$. For example, for $B (\psi, \alpha, \theta)$,
\begin{equation} \label{eq:Taylorexample}
B \simeq B_0 + \delta B_0,
\end{equation}
where $B_0 = B (\psi_0, \alpha_0, \theta)$ is the function $B$ evaluated on the magnetic field line $\psi = \psi_0$ and $\alpha = \alpha_0$, and the operator $\delta$ is 
\begin{equation}
\delta = (\psi - \psi_0) \frac{\partial}{\partial \psi} + (\alpha - \alpha_0) \frac{\partial}{\partial \alpha}.
\end{equation}
Note that we have only Taylor expanded the dependence of $B$ in the directions perpendicular to the magnetic field because the dependence of $B$ along the magnetic field is needed to solve the lowest order equations \eq{eq:FPfluxtube0} and \eq{eq:QNfluxtube0}. By writing equations \eq{eq:FPequation} and \eq{eq:QNequation} using the coordinates $\{ \psi, \alpha, \theta \}$, Taylor expanding all the coefficients as shown in \eq{eq:Taylorexample}, and employing 
\begin{equation} \label{eq:phiaveapprox2}
\langle \phi^\tb \rangle \simeq \langle \phi^\tb \rangle_0 + \langle \delta \rhobf_0 \cdot \nabla \phi^\tb \rangle_0,
\end{equation}
where the average $\langle \ldots \rangle_0$ of a function of space $g(\boldr, t)$ is
\begin{equation}
\langle g \rangle_0 = \frac{1}{2\pi} \int_0^{2\pi} g ( \bR + \rhobf_0, t )\, \dd \varphi,
\end{equation}
we find
\begin{eqnarray} \label{eq:FPlocal}
\fl \frac{\partial h_s^\tb}{\partial t} + \left ( v_{||} (\bun \cdot \nabla_\bR \theta)_0 \frac{\partial}{\partial \theta}  - \mu (\bun \cdot \nabla_\bR B)_0 \frac{\partial}{\partial v_{||}} \right )  h_s^\tb \nonumber\\ + \left ( (\bv_{Ms} \cdot \nabla_\bR \psi)_0 \frac{\partial}{\partial \psi} + \left ( - c \frac{\partial \phi^\lw}{\partial \psi} + \bv_{Ms} \cdot \nabla_\bR \alpha \right )_0 \frac{\partial}{\partial \alpha}\right ) h_s^\tb \nonumber\\ - c \left ( \frac{\partial \langle \phi^\tb \rangle_0}{\partial \psi} \frac{\partial h_s^\tb}{\partial \alpha} - \frac{\partial \langle \phi^\tb \rangle_0}{\partial \alpha} \frac{\partial h_s^\tb}{\partial \psi}\right ) \nonumber\\ + \Bigg ( \Bigg [ \frac{1}{n_s} \frac{\partial n_s}{\partial \psi} + \frac{Z_s e}{T_s} \frac{\partial \phi^\lw}{\partial \psi} + \left ( \frac{m_s (v_{||}^2 + 2\mu B)}{2T_s} - \frac{3}{2} \right ) \frac{1}{T_s} \frac{\partial T_s}{\partial \psi} \Bigg ] f_{Ms} \Bigg )_0 \nonumber\\ \times  c \frac{\partial \langle \phi^\tb \rangle_0}{\partial \alpha} - \left ( \frac{Z_s e}{T_s} f_{Ms} \right )_0 \frac{\partial \langle \phi^\tb \rangle_0}{\partial t} = \nonumber\\ \left ( \left [ \frac{c}{B} (\nabla_\bR \phi^\lw \times \bun) - \bv_{Ms} \right ]\cdot \nabla_\bR \theta \right )_0 \frac{\partial h_s^\tb}{\partial \theta} \nonumber\\ + \left( \frac{c}{B} (\nabla_\bR \theta \times \bun) \cdot \nabla_\bR \psi \right )_0 \left ( \frac{\partial \langle \phi^\tb \rangle_0}{\partial \theta} \frac{\partial h_s^\tb}{\partial \psi} - \frac{\partial \langle \phi^\tb \rangle_0}{\partial \psi} \frac{\partial h_s^\tb}{\partial \theta} \right )  \nonumber\\ + \left( \frac{c}{B} (\nabla_\bR \theta \times \bun) \cdot \nabla_\bR \alpha \right )_0 \left ( \frac{\partial \langle \phi^\tb \rangle_0}{\partial \theta} \frac{\partial h_s^\tb}{\partial \alpha} - \frac{\partial \langle \phi^\tb \rangle_0}{\partial \alpha} \frac{\partial h_s^\tb}{\partial \theta} \right ) \nonumber\\ + \Bigg ( \Bigg [ \frac{1}{n_s} \frac{\partial n_s}{\partial \psi} + \frac{Z_s e}{T_s} \frac{\partial \phi^\lw}{\partial \psi} + \left ( \frac{m_s (v_{||}^2 + 2\mu B)}{2T_s} - \frac{3}{2} \right ) \frac{1}{T_s} \frac{\partial T_s}{\partial \psi} \Bigg ] \nonumber \\ \times \frac{c}{B} (\nabla_\bR \theta \times \bun) \cdot \nabla_\bR \psi \, f_{Ms} \Bigg )_0 \frac{\partial \langle \phi^\tb \rangle_0}{\partial \theta} \nonumber\\ + c \left ( \frac{\partial h_s^\tb}{\partial \alpha} \frac{\partial }{\partial \psi} - \frac{\partial h_s^\tb}{\partial \psi} \frac{\partial}{\partial \alpha}\right )\langle \delta \rhobf_0 \cdot \nabla \phi^\tb \rangle_0 \nonumber\\ + \left ( \frac{Z_s e}{T_s} f_{Ms} \right )_0 \frac{\partial}{\partial t} \langle \delta \rhobf_0 \cdot \nabla \phi^\tb \rangle_0\nonumber\\ - \Bigg ( \Bigg [ \frac{1}{n_s} \frac{\partial n_s}{\partial \psi} + \frac{Z_s e}{T_s} \frac{\partial \phi^\lw}{\partial \psi} + \left ( \frac{m_s (v_{||}^2 + 2\mu B)}{2T_s} - \frac{3}{2} \right ) \frac{1}{T_s} \frac{\partial T_s}{\partial \psi} \Bigg ] f_{Ms} \Bigg )_0 \nonumber\\ \times c \frac{\partial}{\partial \alpha} \langle \delta \rhobf_0 \cdot \nabla \phi^\tb \rangle_0 - \left ( v_{||} \delta (\bun \cdot \nabla_\bR \theta)_0 \frac{\partial}{\partial \theta}  - \mu \delta (\bun \cdot \nabla_\bR B)_0 \frac{\partial}{\partial v_{||}} \right )  h_s^\tb \nonumber\\ - \left ( \delta (\bv_{Ms} \cdot \nabla_\bR \psi)_0 \frac{\partial}{\partial \psi} + \delta \left ( - c \frac{\partial \phi^\lw}{\partial \psi} +\bv_{Ms} \cdot \nabla_\bR \alpha \right )_0 \frac{\partial}{\partial \alpha}\right ) h_s^\tb \nonumber\\ - \delta \Bigg ( \Bigg [ \frac{1}{n_s} \frac{\partial n_s}{\partial \psi} + \frac{Z_s e}{T_s} \frac{\partial \phi^\lw}{\partial \psi} + \left ( \frac{m_s (v_{||}^2 + 2\mu B)}{2T_s} - \frac{3}{2} \right ) \frac{1}{T_s} \frac{\partial T_s}{\partial \psi} \Bigg ] f_{Ms} \Bigg )_0 \nonumber\\ \times  c \frac{\partial \langle \phi^\tb \rangle_0}{\partial \alpha} + \delta \left ( \frac{Z_s e}{T_s} f_{Ms} \right )_0 \frac{\partial \langle \phi^\tb \rangle_0}{\partial t}
\end{eqnarray}
and
\begin{eqnarray} \label{eq:QNlocal}
\fl \sum_s Z_s \int B_0\, h_s^\tb ( \boldr - \rhobf_0, v_{||}, \mu, t)\, \dd v_{||}\, \dd \mu\, \dd \varphi - \sum_s \left ( \frac{Z_s^2 n_s e}{T_s} \right )_0  \phi^\tb =  \nonumber\\ - \sum_s Z_s \int \delta B_0\, h_s^\tb ( \boldr - \rhobf_0, v_{||}, \mu, t)\, \dd v_{||}\, \dd \mu\, \dd \varphi \nonumber\\ + \sum_s Z_s \int B_0\, \delta \rhobf_0 \cdot \nabla_\bR h_s^\tb ( \boldr - \rhobf_0, v_{||}, \mu, t)\, \dd v_{||}\, \dd \mu\, \dd \varphi \nonumber\\ + \sum_s \delta \left ( \frac{Z_s^2 n_s e}{T_s} \right )_0  \phi^\tb.
\end{eqnarray}
All the terms of order $l_\bot/L$ and $\rho_\ast$ are on the right side of the equations. We have neglected terms that are higher order. In deriving equations \eq{eq:FPlocal} and \eq{eq:QNlocal}, we have distinguished between the fast derivatives $\partial/\partial \psi$ and $\partial/\partial \alpha$ and the slow derivative $\partial/\partial \theta$. Because most terms that contain $\partial/\partial \theta$ are small, we have not Taylor expanded the coefficients in terms that contain a derivative with respect to $\theta$. The only exception to this rule is the parallel streaming term $v_{||} \bun \cdot \nabla_\bR \theta (\partial/\partial \theta)$ because it is a lowest order term. Finally, in the small terms we have used the lowest order version of \eq{eq:phiaveapprox2}, $\langle \phi^\tb \rangle \simeq \langle \phi^\tb \rangle_0$.

If we neglect the right side of equations \eq{eq:FPlocal} and \eq{eq:QNlocal}, we obtain equations with coefficients that are independent of $\psi$ and $\alpha$, and we can then Fourier analyze them. The resulting equations are the same as the lowest order flux tube equations \eq{eq:FPfluxtube0} and \eq{eq:QNfluxtube0}, proving that global $\delta f$ simulations give the same results as flux tube simulations for $l_\bot/L \ll 1$. We want to show that the solution in \eq{eq:hhatdef} and \eq{eq:phihatdef} is valid to an order higher in $l_\bot/L$. We substitute equations \eq{eq:hhatdef} and \eq{eq:phihatdef} into equations \eq{eq:FPlocal} and \eq{eq:QNlocal}. The resulting equation can be simplified by using the definitions of $H_{s, \psi}^\tb (\bR, v_{||}, \mu, t)$, $\Phi_\psi^\tb (\boldr, t)$, $H_{s, \alpha}^\tb (\bR, v_{||}, \mu, t)$ and $\Phi_\alpha^\tb (\boldr, t)$ in equations \eq{eq:FourierHpsi}-\eq{eq:FourierPhialpha}, and equations \eq{eq:FPgradpsi}-\eq{eq:QNgradalpha} to obtain that $H_{s, \psi}^\tb (\bR, v_{||}, \mu, t)$, $\Phi_\psi^\tb (\boldr, t)$, $H_{s, \alpha}^\tb (\bR, v_{||}, \mu, t)$ and $\Phi_\alpha^\tb (\boldr, t)$ satisfy
\begin{eqnarray} \label{eq:FPpsi}
\fl \frac{\partial H_{s,\psi}^\tb}{\partial t}+ \left ( v_{||} (\bun \cdot \nabla_\bR \theta)_0 \frac{\partial}{\partial \theta} - \mu (\bun \cdot \nabla_\bR B)_0 \frac{\partial}{\partial v_{||}} \right ) H_{s,\psi}^\tb \nonumber\\ + \left ( (\bv_{Ms} \cdot \nabla_\bR \psi)_0 \frac{\partial}{\partial \psi} + \left (- c \frac{\partial \phi^\lw}{\partial \psi} + \bv_{Ms} \cdot \nabla_\bR \alpha \right )_0 \frac{\partial}{\partial \alpha} \right ) H_{s,\psi}^\tb \nonumber\\ - c \left (\frac{\partial \langle \Phi_\psi^\tb \rangle_0}{\partial \psi} \frac{\partial h_s^\tb}{\partial \alpha} - \frac{\partial \langle \Phi_\psi^\tb \rangle_0}{\partial \alpha} \frac{\partial h_s^\tb}{\partial \psi} \right )  \nonumber\\ - c \left (\frac{\partial \langle \phi^\tb \rangle_0}{\partial \psi} \frac{\partial H_{s,\psi}^\tb}{\partial \alpha} - \frac{\partial \langle \phi^\tb \rangle_0}{\partial \alpha} \frac{\partial H_{s,\psi}^\tb}{\partial \psi} \right ) \nonumber\\ + \Bigg ( \Bigg [ \frac{1}{n_s} \frac{\partial n_s}{\partial \psi} + \frac{Z_s e}{T_s} \frac{\partial \phi^\lw}{\partial \psi} + \left ( \frac{m_s (v_{||}^2 + 2\mu B)}{2T_s} - \frac{3}{2} \right ) \frac{1}{T_s} \frac{\partial T_s}{\partial \psi} \Bigg ] f_{Ms} \Bigg )_0 \nonumber\\ \times c \frac{\partial \langle \Phi_\psi^\tb \rangle_0}{\partial \alpha} - \left ( \frac{Z_s e}{T_s} f_{Ms} \right )_0 \frac{\partial \langle \Phi_\psi^\tb \rangle_0}{\partial t} = - v_{||} \frac{\partial}{\partial \psi} (\bun \cdot \nabla_\bR \theta)_0 \frac{\partial\underline{h}_s^\tb}{\partial \theta} \nonumber\\ + \mu \frac{\partial}{\partial \psi} ( \bun \cdot \nabla_\bR B )_0 \frac{\partial \underline{h}_s^\tb}{\partial v_{||}} -  \Bigg ( \frac{\partial}{\partial \psi} (\bv_{Ms} \cdot \nabla_\bR \psi)_0 \frac{\partial}{\partial \psi} \nonumber\\ + \frac{\partial}{\partial \psi} \left (- c \frac{\partial \phi^\lw}{\partial \psi} + \bv_{Ms} \cdot \nabla_\bR \alpha \right )_0 \frac{\partial}{\partial \alpha} \Bigg ) \underline{h}_s^\tb \nonumber\\- \frac{\partial}{\partial \psi} \Bigg ( \Bigg [ \frac{1}{n_s} \frac{\partial n_s}{\partial \psi} + \frac{Z_s e}{T_s} \frac{\partial \phi^\lw}{\partial \psi} + \left ( \frac{m_s (v_{||}^2 + 2\mu B)}{2T_s} - \frac{3}{2} \right ) \frac{1}{T_s} \frac{\partial T_s}{\partial \psi} \Bigg ] f_{Ms} \Bigg )_0 \nonumber \\ \times c \frac{\partial \langle \phi^\tb \rangle_0}{\partial \alpha} + \frac{\partial}{\partial \psi} \left ( \frac{Z_s e}{T_s} f_{Ms} \right )_0 \frac{\partial \langle \phi^\tb \rangle_0}{\partial t} \nonumber\\ + c \left (\frac{\partial h_s^\tb}{\partial \alpha} \frac{\partial}{\partial \psi} - \frac{\partial h_s^\tb}{\partial \psi} \frac{\partial}{\partial \alpha} \right ) \left \langle \frac{\partial \rhobf_0}{\partial \psi} \cdot \nabla \phi^\tb \right \rangle_0 \nonumber\\ - \Bigg ( \Bigg [ \frac{1}{n_s} \frac{\partial n_s}{\partial \psi} + \frac{Z_s e}{T_s} \frac{\partial \phi^\lw}{\partial \psi} + \left ( \frac{m_s (v_{||}^2 + 2\mu B)}{2T_s} - \frac{3}{2} \right ) \frac{1}{T_s} \frac{\partial T_s}{\partial \psi} \Bigg ] f_{Ms} \Bigg )_0 \nonumber\\ \times c \frac{\partial}{\partial \alpha}  \left \langle \frac{\partial \rhobf_0}{\partial \psi} \cdot \nabla \phi^\tb \right \rangle_0 + \left ( \frac{Z_s e}{T_s} f_{Ms} \right ) \frac{\partial}{\partial t} \left \langle \frac{\partial \rhobf_0}{\partial \psi} \cdot \nabla \phi^\tb \right \rangle_0,
\end{eqnarray}
\begin{eqnarray} \label{eq:QNpsi}
\fl \sum_s Z_s \int B_0\, H_{s,\psi}^\tb ( \boldr - \rhobf_0, v_{||}, \mu, t)\, \dd v_{||}\, \dd \mu\, \dd \varphi - \sum_s \left ( \frac{Z_s^2 n_s e}{T_s} \right )_0  \Phi_\psi^\tb =  \nonumber\\ - \sum_s Z_s \int \frac{\partial B_0}{\partial \psi}\, h_s^\tb ( \boldr - \rhobf_0, v_{||}, \mu, t)\, \dd v_{||}\, \dd \mu\, \dd \varphi \nonumber\\ + \sum_s Z_s \int B_0\, \frac{\partial \rhobf_0}{\partial \psi} \cdot \nabla_\bR h_s^\tb ( \boldr - \rhobf_0, v_{||}, \mu, t)\, \dd v_{||}\, \dd \mu\, \dd \varphi \nonumber\\ + \sum_s \frac{\partial}{\partial \psi} \left ( \frac{Z_s^2 n_s e}{T_s} \right )_0  \phi^\tb,
\end{eqnarray}
\begin{eqnarray} \label{eq:FPalpha}
\fl \frac{\partial H_{s,\alpha}^\tb}{\partial t}+ \left ( v_{||} (\bun \cdot \nabla_\bR \theta)_0 \frac{\partial}{\partial \theta} - \mu (\bun \cdot \nabla_\bR B)_0 \frac{\partial}{\partial v_{||}} \right ) H_{s,\alpha}^\tb \nonumber\\ + \left ( (\bv_{Ms} \cdot \nabla_\bR \psi)_0 \frac{\partial}{\partial \psi} + \left (- c \frac{\partial \phi^\lw}{\partial \psi} + \bv_{Ms} \cdot \nabla_\bR \alpha \right )_0 \frac{\partial}{\partial \alpha} \right ) H_{s,\alpha}^\tb \nonumber\\ - c \left (\frac{\partial \langle \Phi_\alpha^\tb \rangle_0}{\partial \psi} \frac{\partial h_s^\tb}{\partial \alpha} - \frac{\partial \langle \Phi_\alpha^\tb \rangle_0}{\partial \alpha} \frac{\partial h_s^\tb}{\partial \psi} \right )  \nonumber\\ - c \left (\frac{\partial \langle \phi^\tb \rangle_0}{\partial \psi} \frac{\partial H_{s,\alpha}^\tb}{\partial \alpha} - \frac{\partial \langle \phi^\tb \rangle_0}{\partial \alpha} \frac{\partial H_{s,\alpha}^\tb}{\partial \psi} \right ) \nonumber\\ + \Bigg ( \Bigg [ \frac{1}{n_s} \frac{\partial n_s}{\partial \psi} + \frac{Z_s e}{T_s} \frac{\partial \phi^\lw}{\partial \psi} + \left ( \frac{m_s (v_{||}^2 + 2\mu B)}{2T_s} - \frac{3}{2} \right ) \frac{1}{T_s} \frac{\partial T_s}{\partial \psi} \Bigg ] f_{Ms} \Bigg )_0 \nonumber\\ \times c \frac{\partial \langle \Phi_\alpha^\tb \rangle_0}{\partial \alpha} - \left ( \frac{Z_s e}{T_s} f_{Ms} \right )_0 \frac{\partial \langle \Phi_\alpha^\tb \rangle_0}{\partial t} = - v_{||} \frac{\partial}{\partial \alpha} (\bun \cdot \nabla_\bR \theta)_0 \frac{\partial\underline{h}_s^\tb}{\partial \theta} \nonumber\\ + \mu \frac{\partial}{\partial \alpha} ( \bun \cdot \nabla_\bR B )_0 \frac{\partial \underline{h}_s^\tb}{\partial v_{||}} -  \Bigg ( \frac{\partial}{\partial \alpha} (\bv_{Ms} \cdot \nabla_\bR \psi)_0 \frac{\partial}{\partial \psi} \nonumber\\ + \frac{\partial}{\partial \alpha} \left (\bv_{Ms} \cdot \nabla_\bR \alpha \right )_0 \frac{\partial}{\partial \alpha} \Bigg ) \underline{h}_s^\tb \nonumber\\- \frac{\partial}{\partial \alpha} \Bigg ( \Bigg [ \frac{1}{n_s} \frac{\partial n_s}{\partial \psi} + \frac{Z_s e}{T_s} \frac{\partial \phi^\lw}{\partial \psi} + \left ( \frac{m_s (v_{||}^2 + 2\mu B)}{2T_s} - \frac{3}{2} \right ) \frac{1}{T_s} \frac{\partial T_s}{\partial \psi} \Bigg ] f_{Ms} \Bigg )_0 \nonumber\\ \times c \frac{\partial \langle \phi^\tb \rangle_0}{\partial \alpha} + \frac{\partial}{\partial \alpha} \left ( \frac{Z_s e}{T_s} f_{Ms} \right )_0 \frac{\partial \langle \phi^\tb \rangle_0}{\partial t} \nonumber\\ + c \left (\frac{\partial h_s^\tb}{\partial \alpha} \frac{\partial}{\partial \psi} - \frac{\partial h_s^\tb}{\partial \psi} \frac{\partial}{\partial \alpha} \right ) \left \langle \frac{\partial \rhobf_0}{\partial \alpha} \cdot \nabla \phi^\tb \right \rangle_0 \nonumber\\ - \Bigg ( \Bigg [ \frac{1}{n_s} \frac{\partial n_s}{\partial \psi} + \frac{Z_s e}{T_s} \frac{\partial \phi^\lw}{\partial \psi} + \left ( \frac{m_s (v_{||}^2 + 2\mu B)}{2T_s} - \frac{3}{2} \right ) \frac{1}{T_s} \frac{\partial T_s}{\partial \psi} \Bigg ] f_{Ms} \Bigg )_0 \nonumber\\ \times c \frac{\partial}{\partial \alpha}  \left \langle \frac{\partial \rhobf_0}{\partial \alpha} \cdot \nabla \phi^\tb \right \rangle_0 + \left ( \frac{Z_s e}{T_s} f_{Ms} \right ) \frac{\partial}{\partial t} \left \langle \frac{\partial \rhobf_0}{\partial \alpha} \cdot \nabla \phi^\tb \right \rangle_0
\end{eqnarray}
and
\begin{eqnarray} \label{eq:QNalpha}
\fl \sum_s Z_s \int B_0\, H_{s,\alpha}^\tb ( \boldr - \rhobf_0, v_{||}, \mu, t)\, \dd v_{||}\, \dd \mu\, \dd \varphi - \sum_s \left ( \frac{Z_s^2 n_s e}{T_s} \right )_0  \Phi_\alpha^\tb =  \nonumber\\ - \sum_s Z_s \int \frac{\partial B_0}{\partial \alpha}\, h_s^\tb ( \boldr - \rhobf_0, v_{||}, \mu, t)\, \dd v_{||}\, \dd \mu\, \dd \varphi \nonumber\\ + \sum_s Z_s \int B_0\, \frac{\partial \rhobf_0}{\partial \alpha} \cdot \nabla_\bR h_s^\tb ( \boldr - \rhobf_0, v_{||}, \mu, t)\, \dd v_{||}\, \dd \mu\, \dd \varphi.
\end{eqnarray}

Substituting \eq{eq:hhatdef} and \eq{eq:phihatdef} into \eq{eq:FPlocal} and \eq{eq:QNlocal}, using equations \eq{eq:FPpsi}-\eq{eq:QNalpha}, and employing
\begin{eqnarray}
\fl h_s^\tb ( \boldr - \rhobf_0, v_{||}, \mu, t) \simeq \hat{h}_s^\tb ( \boldr - \rhobf_0, v_{||}, \mu, t) \nonumber\\ + (\psi(\boldr) - \psi_0 - (\rhobf \cdot \nabla \psi)_0 ) H_{s,\psi}^\tb ( \boldr - \rhobf_0, v_{||}, \mu, t) \nonumber\\ + (\alpha(\boldr) - \alpha_0 - (\rhobf \cdot \nabla \alpha)_0 ) H_{s,\alpha}^\tb ( \boldr - \rhobf_0, v_{||}, \mu, t),
\end{eqnarray}
\begin{eqnarray}
\fl \langle \phi^\tb \rangle_0 \simeq \langle \hat{\phi}^\tb \rangle_0 + (\psi(\bR) - \psi_0) \langle \Phi_\psi^\tb \rangle_0 + (\alpha(\bR) - \alpha_0 ) \langle \Phi_\alpha^\tb \rangle_0 \nonumber\\ + \langle (\rhobf \cdot \nabla_\bR \psi)_0 \Phi_\psi^\tb \rangle_0 + \langle (\rhobf \cdot \nabla_\bR \alpha)_0 \Phi_\alpha^\tb \rangle_0
\end{eqnarray}
and
\begin{eqnarray}
\fl \langle \delta \rhobf_0 \cdot \nabla \phi^\tb \rangle_0 \simeq (\psi(\bR) - \psi_0) \left \langle \frac{\partial \rhobf_0}{\partial \psi} \cdot \nabla \hat{\phi}^\tb \right \rangle_0 + (\alpha(\bR) - \alpha_0) \left \langle \frac{\partial \rhobf_0}{\partial \alpha} \cdot \nabla \hat{\phi}^\tb \right \rangle_0,
\end{eqnarray}
we find
\begin{eqnarray} \label{eq:FPlocalfinal}
\fl \frac{\partial \hat{h}_s^\tb}{\partial t} + \left ( v_{||} (\bun \cdot \nabla_\bR \theta)_0 \frac{\partial}{\partial \theta}  - \mu (\bun \cdot \nabla_\bR B)_0 \frac{\partial}{\partial v_{||}} \right )  \hat{h}_s^\tb \nonumber\\ + \left ( (\bv_{Ms} \cdot \nabla_\bR \psi)_0 \frac{\partial}{\partial \psi} + \left ( - c \frac{\partial \phi^\lw}{\partial \psi} + \bv_{Ms} \cdot \nabla_\bR \alpha \right )_0 \frac{\partial}{\partial \alpha}\right ) \hat{h}_s^\tb \nonumber\\ - c \left ( \frac{\partial \langle \hat{\phi}^\tb \rangle_0}{\partial \psi} \frac{\partial \hat{h}_s^\tb}{\partial \alpha} - \frac{\partial \langle \hat{\phi}^\tb \rangle_0}{\partial \alpha} \frac{\partial \hat{h}_s^\tb}{\partial \psi}\right ) \nonumber\\ + \Bigg ( \Bigg [ \frac{1}{n_s} \frac{\partial n_s}{\partial \psi} + \frac{Z_s e}{T_s} \frac{\partial \phi^\lw}{\partial \psi} + \left ( \frac{m_s (v_{||}^2 + 2\mu B)}{2T_s} - \frac{3}{2} \right ) \frac{1}{T_s} \frac{\partial T_s}{\partial \psi} \Bigg ] f_{Ms} \Bigg )_0 \nonumber\\ \times c \frac{\partial \langle \hat{\phi}^\tb \rangle_0}{\partial \alpha} - \left ( \frac{Z_s e}{T_s} f_{Ms} \right )_0 \frac{\partial \langle \hat{\phi}^\tb \rangle_0}{\partial t} = - (\bv_{Ms} \cdot \nabla_\bR \psi)_0 H_{s,\psi}^\tb \nonumber\\ + \left ( c \frac{\partial \phi^\lw}{\partial \psi} - \bv_{Ms} \cdot \nabla_\bR \alpha \right )_0 H_{s,\alpha}^\tb \nonumber\\ + \left ( \left [ \frac{c}{B} (\nabla_\bR \phi^\lw \times \bun) - \bv_{Ms} \right ]\cdot \nabla_\bR \theta \right )_0 \frac{\partial h_s^\tb}{\partial \theta}\nonumber\\ + c \left ( \frac{\partial \langle \hat{\phi}^\tb \rangle_0}{\partial \psi} H_{s, \alpha}^\tb - \frac{\partial \langle \hat{\phi}^\tb \rangle_0}{\partial \alpha} H_{s,\psi}^\tb \right )\nonumber\\ + c \Bigg ( \Bigg \langle \Phi_\psi^\tb + \frac{\partial \rhobf_0}{\partial \psi} \cdot \nabla \hat{\phi}^\tb + ( \rhobf \cdot \nabla_\bR \psi )_0 \frac{\partial \Phi_\psi^\tb}{\partial \psi} \nonumber\\ + (\rhobf \cdot \nabla_\bR \alpha)_0 \frac{\partial \Phi_\alpha^\tb}{\partial \psi} \Bigg \rangle_0 \frac{\partial \hat{h}_s^\tb}{\partial \alpha} - \Bigg \langle \Phi_\alpha^\tb + \frac{\partial \rhobf_0}{\partial \alpha} \cdot \nabla \hat{\phi}^\tb  \nonumber\\ + ( \rhobf \cdot \nabla_\bR \psi )_0 \frac{\partial \Phi_\psi^\tb}{\partial \alpha} + (\rhobf \cdot \nabla_\bR \alpha)_0 \frac{\partial \Phi_\alpha^\tb}{\partial \alpha} \Bigg \rangle_0 \frac{\partial \hat{h}_s^\tb}{\partial \psi}\Bigg ) \nonumber\\ + \left( \frac{c}{B} (\nabla_\bR \theta \times \bun) \cdot \nabla_\bR \psi \right )_0 \left ( \frac{\partial \langle \hat{\phi}^\tb \rangle_0}{\partial \theta} \frac{\partial \hat{h}_s^\tb}{\partial \psi} - \frac{\partial \langle \hat{\phi}^\tb \rangle_0}{\partial \psi} \frac{\partial \hat{h}_s^\tb}{\partial \theta} \right )  \nonumber\\ + \left( \frac{c}{B} (\nabla_\bR \theta \times \bun) \cdot \nabla_\bR \alpha \right )_0 \left ( \frac{\partial \langle \hat{\phi}^\tb \rangle_0}{\partial \theta} \frac{\partial \hat{h}_s^\tb}{\partial \alpha} - \frac{\partial \langle \hat{\phi}^\tb \rangle_0}{\partial \alpha} \frac{\partial \hat{h}_s^\tb}{\partial \theta} \right ) \nonumber\\ -\Bigg ( \Bigg [ \frac{1}{n_s} \frac{\partial n_s}{\partial \psi} + \frac{Z_s e}{T_s} \frac{\partial \phi^\lw}{\partial \psi} + \left ( \frac{m_s (v_{||}^2 + 2\mu B)}{2T_s} - \frac{3}{2} \right ) \frac{1}{T_s} \frac{\partial T_s}{\partial \psi} \Bigg ] f_{Ms} \Bigg )_0 \nonumber\\ \times c \left \langle \Phi_\alpha^\tb + \frac{\partial \rhobf_0}{\partial \alpha} \cdot \nabla \hat{\phi}^\tb + ( \rhobf_0 \cdot \nabla_\bR \psi )_0 \frac{\partial \Phi_\psi^\tb}{\partial \alpha} + (\rhobf \cdot \nabla_\bR \alpha)_0 \frac{\partial \Phi_\alpha^\tb}{\partial \alpha} \right \rangle_0 \nonumber\\ + \Bigg ( \Bigg [ \frac{1}{n_s} \frac{\partial n_s}{\partial \psi} + \frac{Z_s e}{T_s} \frac{\partial \phi^\lw}{\partial \psi} + \left ( \frac{m_s (v_{||}^2 + 2\mu B)}{2T_s} - \frac{3}{2} \right ) \frac{1}{T_s} \frac{\partial T_s}{\partial \psi} \Bigg ] \nonumber \\ \times \frac{c}{B} (\nabla_\bR \theta \times \bun) \cdot \nabla_\bR \psi \, f_{Ms} \Bigg )_0 \frac{\partial \langle \hat{\phi}^\tb \rangle_0}{\partial \theta}
\end{eqnarray}
and
\begin{eqnarray} \label{eq:QNlocalfinal}
\fl \sum_s Z_s \int B_0\, \hat{h}_s^\tb ( \boldr - \rhobf_0, v_{||}, \mu, t)\, \dd v_{||}\, \dd \mu\, \dd \varphi - \sum_s \left ( \frac{Z_s^2 n_s e}{T_s} \right )_0  \hat{\phi}^\tb = \nonumber\\  \sum_s Z_s \int B_0 \Big [ (\rhobf \cdot \nabla \psi)_0 H_{s,\psi}^\tb ( \boldr - \rhobf_0, v_{||}, \mu, t) \nonumber\\ + (\rhobf \cdot \nabla \alpha)_0 H_{s,\alpha}^\tb ( \boldr - \rhobf_0, v_{||}, \mu, t) \Big ] \dd v_{||}\, \dd \mu\, \dd \varphi.
\end{eqnarray}
To obtain these equations, we have used that the coefficients of the Fourier series for $\hat{h}_s^\tb$, $\hat{\phi}^\tb$, $H_{s, \psi}^\tb$, $\Phi_\psi^\tb$, $H_{s, \alpha}^\tb$ and $\Phi_\alpha^\tb$, given in \eq{eq:Fourierhhat}-\eq{eq:FourierPhialpha}, do not depend on $\psi$ or $\alpha$ because they are evaluated at the fixed values $\psi = \psi_0$ and $\alpha = \alpha_0$.

Equations \eq{eq:FPlocalfinal} and \eq{eq:QNlocalfinal} vanish because they are the Fourier transforms of equations \eq{eq:FPfluxtube} and \eq{eq:QNfluxtube}. To show that equations \eq{eq:FPlocalfinal} and \eq{eq:QNlocalfinal} are the Fourier transforms of equations \eq{eq:FPfluxtube} and \eq{eq:QNfluxtube}, we use equations \eq{eq:Fourierhhat} and \eq{eq:Fourierphihat} to find
\begin{eqnarray} \label{eq:phiaveapprox3}
\fl \langle \hat{\phi}^\tb \rangle_0 \simeq \sum_{k_\psi, k_\alpha} \Bigg ( \underline{\phi}^\tb J_0 (\Lambda_s) + \frac{2J_1(\Lambda_s)}{\Lambda_s}\frac{i \mu B}{2 \Omega_s^2} \left ( \matI - \bun \bun \right ) : \nabla_\bR \bk_\bot\, \underline{\phi}^\tb \nonumber\\ - G(\Lambda_s) \frac{i k_\bot \mu^2 B^2 }{4\Omega_s^4 }  \bk_\bot \cdot \nabla_\bR k_\bot\, \underline{\phi}^\tb \Bigg ) \exp ( i k_\psi \psi (\bR) + i k_\alpha \alpha (\bR)),
\end{eqnarray}
and
\begin{eqnarray} \label{eq:hrvapprox3}
\fl \int_0^{2\pi} \hat{h}_s^\tb (\boldr - \rhobf_0, v_{||}, \mu, t)\, \dd \varphi = 2\pi \sum_{k_\psi, k_\alpha} \Bigg ( \underline{h}_s^\tb J_0 (\lambda_s) + \frac{2J_1(\lambda_s)}{\lambda_s}\frac{i \mu B}{2 \Omega_s^2} \left ( \matI - \bun \bun \right ) : \nabla_\bR \bk_\bot\, \underline{h}_s^\tb \nonumber\\ - G(\lambda_s) \frac{i k_\bot \mu^2 B^2 }{4\Omega_s^4 }  \bk_\bot \cdot \nabla_\bR k_\bot\, \underline{h}_s^\tb \Bigg ) \exp ( i k_\psi \psi(\boldr) + i k_\alpha (\boldr) ).
\end{eqnarray}
Note the difference between these results and equations \eq{eq:phiaveapprox} and \eq{eq:hrvapprox}. 

\section*{References}

\end{document}